\documentclass[%
superscriptaddress,
 preprint,
 amsmath,amssymb,
 aps,
]{revtex4-2}

\usepackage{amsmath,amssymb,amsfonts}%

\usepackage{graphicx}
\usepackage{dcolumn}
\usepackage{bm}
\usepackage{xcolor}%

\newcommand{\sub}[1]{$_{\mathrm {#1}}$}
\newcommand{\subm}[1]{_{\mathrm {#1}}}

\newcommand{\TN}{T_{\mathrm{N}}}
\newcommand{\thetaK}{\theta_{\mathrm{K}}}
\newcommand{\thetaT}{\theta_{\mathrm{tot}}}
\newcommand{\phim}{\phi_{\mathrm{m}}}

\newcommand{\rev}[1]{#1}

\begin{document}


\title{Altermagnetic domains in MnTe\\ with large spontaneous magneto-optical Kerr rotation}

\author{Gakuto Watanabe}
\affiliation{%
 Department of Electrical, Electronic, and Digital Science and Engineering, Graduate School of Engineering, Kyoto University, Kyoto 615-8510, Kyoto, Japan}

\author{Soichiro Yamane}%
\affiliation{%
 Department of Electrical, Electronic, and Digital Science and Engineering, Graduate School of Engineering, Kyoto University, Kyoto 615-8510, Kyoto, Japan}

\author{Ryotaro Maki}
\affiliation{%
 Department of Electrical, Electronic, and Digital Science and Engineering, Graduate School of Engineering, Kyoto University, Kyoto 615-8510, Kyoto, Japan}

\author{Atsutoshi Ikeda}
\affiliation{%
 Department of Electrical, Electronic, and Digital Science and Engineering, Graduate School of Engineering, Kyoto University, Kyoto 615-8510, Kyoto, Japan}

\author{Akimitsu Kirikoshi}
\affiliation{Research Institute for Interdisciplinary Science, Okayama University, Okayama, 700-8530, Okayama, Japan}

\author{Junya Otsuki}
\affiliation{Research Institute for Interdisciplinary Science, Okayama University, Okayama, 700-8530, Okayama, Japan}

\author{Takuya Aoyama}
\affiliation{Department of Physics, Graduate School of Science, Tohoku University, Sendai, 980-8578, Miyagi, Japan}
\affiliation{Department of Physics, Graduate School of Advanced Science and Engineering, Hiroshima University, Higashi-Hiroshima, 739-8530, Hiroshima, Japan}

\affiliation{International Institute for Sustainability with Knotted Chiral Meta Matter (WPI-SKCM2), Hiroshima University, Higashi-Hiroshima, 739-8526, Hiroshima, Japan}

\author{Kenya Ohgushi}

\author{Shingo Yonezawa}
\email{yonezawa@sk.kuee.kyoto-u.ac.jp}
\affiliation{%
 Department of Electrical, Electronic, and Digital Science and Engineering, Graduate School of Engineering, Kyoto University, Kyoto 615-8510, Kyoto, Japan}

\

\date{\today}

\begin{abstract}
Altermagnets are a new class of magnets accompanying global time-reversal symmetry breaking (TRSB) without net magnetization, candidate materials for next-generation high-density magnetic devices.
The TRSB should results in formation of novel altermagnetic domains.
\rev{Properties of such domains} are essentially important \rev{both for fundamental understanding of the altermagnetic order as well as} for future device application, but yet unexplored. 
Here, we report comprehensive magneto-optical Kerr effect (MOKE) study on MnTe.
With scanning MOKE microscopy, we succeeded in visualizing bulk altermagnetic domains.
\rev{The domain configuration is largely altered  by thermal cycles, showing spontaneous nature of the altermagnetic order. 
We also clearly show that the altermagnetic domains can be polled by external $c$-axis magnetic field, demonstrating that the altermagnetic order with negligible net magnetic moment indeed can couple with external magnetic fields.}
Furthermore, we revealed that the Kerr rotation of MnTe reaches 10~mrad at low temperatures, a large response exceeding those in standard ferromagnets in spite of million times smaller bulk magnetization. 
\rev{Theoretical calculation shows that this large Kerr rotation originates intrinsically from the altermagnetic order of the Mn magnetic moment. 
Microscopically, the MOKE response stems from large non-zero optical Hall contribution of optical transitions around the $L$ points exhibiting avoided cancellation due to the breaking of the mirror symmetry.
This work provide important bases on altermagnetism both for fundamental as well as application point of views, placing MnTe as a promising material toward understanding altermagnetic domain behavior and realization of altermagnetic devices capable of high-density assemblies.}
\end{abstract}

\maketitle

\section{Introduction}

Ordered states in materials with spontaneous symmetry breakings lead to various functionalities, as exemplified by ferromagnetism and ferroelectricity. 
Physical responses originating from such orders have been central working principles of various devices.
Moreover, broken symmetries often result in formation of domains having distinct symmetry characteristics, such as ferromagnetic domains having magnetic moments pointing to different orientations~\cite{ChikazumiText}.
Properties of such domains are indispensable toward future device applications.
Finding of ordered states exhibiting large responses, as well as visualization of domain controllability, are fundamental and crucial steps in the study of novel functional materials.

Ferromagnets have macroscopic magnetization and broken global time-reversal symmetry (TRS), whereas antiferromagnets ordinarily have no net magnetization and preserve TRS. 
However, some antiferromagnets with special combinations of crystal and magnetic structures have been known to show broken TRS~\cite{TAVGER1956_1954submitsion, borovik1960piezomagnetism,Solovyev1997.PhysRevB.55.8060,Ahn2019.PhysRevB.99.184432,Noda2016.PhysChemChemPhys.18.13294,Naka2019.NatureCommun.10.4305,Hayami2019.JPhysSocJpn.88.123703,Smejkal2020.SciAdv.6.aaz8809}. 
A class of such TRS-broken collinear antiferromagnets is recently gaining tremendous attention as ``altermagnets''~\cite{Smejkal2022.PhysRevX.12.031042,Smejkal2022.PhysRevX.12.040501,Mazin2022.PhysRevX.12.040002}, which are characterized by global TRS breaking (TRSB) accompanying wavenumber $k$ dependent spin splitting in their electronic bands~\cite{Noda2016.PhysChemChemPhys.18.13294,Ahn2019.PhysRevB.99.184432,Naka2019.NatureCommun.10.4305,Hayami2019.JPhysSocJpn.88.123703,Smejkal2020.SciAdv.6.aaz8809}.
These unique features of altermagnets can be utilized for efficient high-density magnetic devices free from stray-field interferences~\cite{Naka2019.NatureCommun.10.4305,Smejkal2022.PhysRevX.12.031042,Smejkal2022.PhysRevX.12.040501}.

\begin{figure} 
	\centering
	\includegraphics[width=1\textwidth]{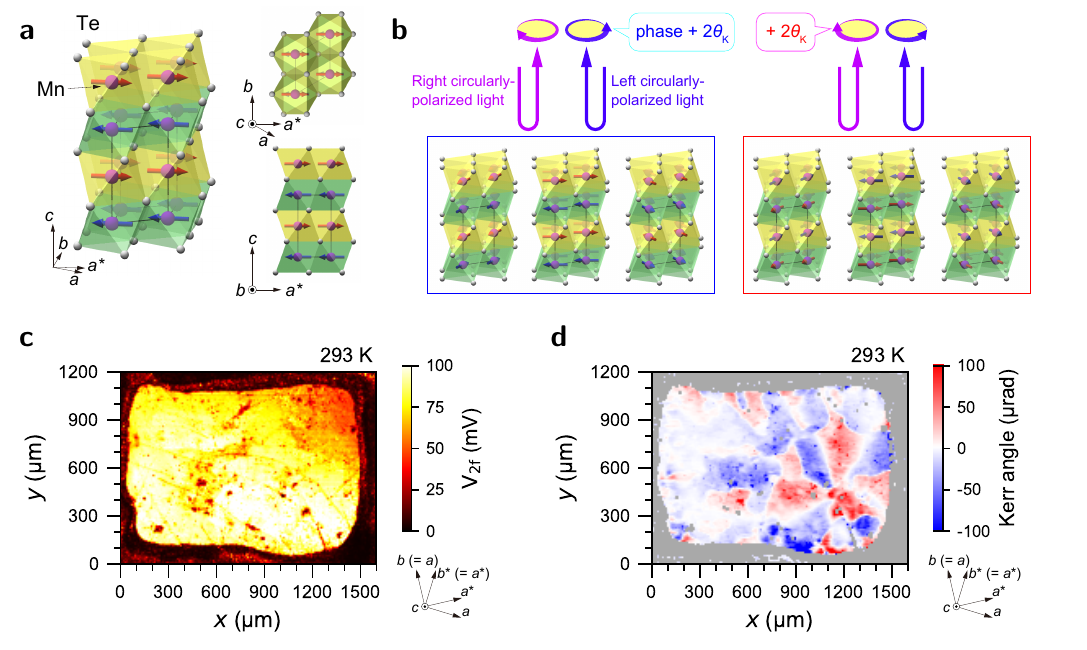} 

	\caption{\textbf{Altermagnetic order and magneto-optical imaging of MnTe.}
 	\textbf{a}. Crystal and magnetic structure of the altermagnet MnTe.
     The crystal structure is drawn based on Ref.~\cite{DSa2005.JMagMagMater.285.267}. 
     The blue and red arrows indicate magnetic moments on Mn ions, pointing $+a^\ast$ and $-a^\ast$ directions, respectively~\cite{Aoyama2024.PhysRevMaterials.8.L041402}.
     The two distinct Mn layers are distinguished by the color of surrounding Te octahedra (yellow or green).
    \textbf{b}. Six possible magnetic ordered structures and illustration of their MOKE responses. 
      Among the six possible magnetic structures, three structures that are mutually converted by $\pm 120^\circ$ rotations exhibit Kerr rotation of positive sign, whereas the other three exhibit negative Kerr rotation.
      Kerr rotation are measured as the excessive phase factor gained by either of the two circular polarized lights.
	\textbf{c}. 6-$\mu$m-step mapping of the 2nd harmonic voltage $V_{2f}$, which is proportional to the reflected light power.
       The bright region is the MnTe sample.
       Directions of crystalline axes are indicated.
    \textbf{d}. Kerr-angle mapping of the MnTe sample measured with  6-$\mu$m steps. 
      The red and blue colors indicate the size and sign of the Kerr angle, as indicated by the color bar.
      We succeeded in observing clear magnetic domains with positive or negative Kerr angles.
      The pixels with $V_{2f} < 30$~mV are shown with gray, since these pixels have very weak light reflection and the Kerr angle contains large uncertainty.   
       }
	\label{fig:fig1} 
\end{figure}

MnTe is one of the most promising candidate for an altermagnet~\cite{Smejkal2022.PhysRevX.12.031042}.
This material form a hexagonal crystal lattice of the space group $P6_3/mmc$, consisting of alternating stacking of triangular layers of Mn and of Te (Fig.~\ref{fig:fig1}{\bf a})~\cite{DSa2005.JMagMagMater.285.267}.
There are two Mn layers with different neighboring Te arrangements, as indicated by the color of the surrounding Te octahedra. 
Notice that Mn sites surrounded by green and yellow octahedra cannot be interchanged by translation or inversion operations. 
Upon the N\'eel order occurring below $\TN\sim 300$ -- 310~K, magnetic moments on the Mn site, $\bm{m}\subm{Mn}$, point an in-plane direction perpendicular to the crystalline $a$ axis, namely the $a^\ast$ direction~\cite{Aoyama2024.PhysRevMaterials.8.L041402}. 
They order ferromagnetically within a Mn layer but antiferromagnetically between neighboring layers.
For this magnetic order with the magnetic point group $m'm'm$, a magnetic structure with $\bm{m}\subm{Mn}$ pointing the $+a^\ast$ direction on the yellow Mn layer cannot be converted only by translation or inversion to its time-reversal counterpart, namely the structure with $\bm{m}\subm{Mn}$ pointing the $-a^\ast$ direction on the yellow Mn layer.
Thus, TRS in MnTe is globally broken, manifesting the altermagnetic nature.
Combined with the hexagonal crystal structure, electronic bands of MnTe are believed to exhibit unconventional $g$-wave type spin splitting in the reciprocal space~\cite{Osumi2024.PhysRevB.109.115102}.
The TRSB and altermagnetic band splittings have been indeed observed using several experimental probes~\cite{Betancourt2023.PhysRevLett.130.036702,Aoyama2024.PhysRevMaterials.8.L041402, Amin2024.Nature.636.348, Takegami2025.PhysRevLett.135.196502, Yamamoto2025.PhysRevAppl.24.034037}.

Due to the global TRSB in altermagnets, altermagnets should form magnetic domains having opposite TRSB characteristics. 
Nevertheless, altermagnetic domains remains unexplored, and detailed and comprehensive studies on domain stability and controllability are quite urgent.
In MnTe, due to its hexagonal crystal structure, there are six possible directions for the in-plane $\bm{m}\subm{Mn}$ and thus there are six different magnetic structures as illustrated in Fig.~\ref{fig:fig1}{\bf b}.
Among them, the structures enclosed in the blue rectangle and those in the red rectangle are time-reversal counterparts.
Thus MnTe below $\TN$ in zero external stimuli should exhibits six types of domains, in which one of the six magnetic structures is realized~\cite{Amin2024.Nature.636.348}.
Indeed, such six types of domains in MnTe have been observed by using high-energy X-ray facilities with a sophisticated combination of X-ray magnetic circular dichroism (XMCD) and X-ray magnetic linear dichroism (XMLD) for 30-nm-thick thin film~\cite{Amin2024.Nature.636.348} and 200-nm-thick single-crystal lamella~\cite{Yamamoto2025.PhysRevAppl.24.034037}.

Magneto-optical Kerr effect (MOKE) is a safe and promising laboratory-level tool to investigate TRSB in MnTe.
The MOKE is a change in the state of light polarization upon reflection from magnetic materials with broken TRS.
Conventionally, MOKE has been considered as an optical probe of magnetization~\cite{BlundellText}.
Recently it is known that MOKE can be observed in systems exhibiting non-trivial TRSB with zero or tiny net magnetization~\cite{Solovyev1997.PhysRevB.55.8060,Higo2018.NaturePhotonics.12.73,Mazin2023.PhysRevB.107.L100418,Farhang2025.arXiv.2510.19709,Yoshimochi2026.arXiv2601.13723}. 
MOKE can be utilized in future as an optical reading method of information stored in such non-trivial TRSB states.
Moreover, MOKE is a local probe allowing observation of real-space distribution of TRSB by a scanning technique.
Compared with the X-ray-based techniques available only in synchrotron facilities~\cite{Amin2024.Nature.636.348,Yamamoto2025.PhysRevAppl.24.034037,Takegami2025.PhysRevLett.135.196502}, scanning MOKE measurements using ordinary optical wavelengths available in ordinary laboratories have various advantages such as compatibilities with various external parameters and wide scan areas. 
We comment that, although MOKE of MnTe using visible light has been recently reported~\cite{Hubert2025.PhysStatSolidiB.262.2400541}, spatially-averaged magnetic-field induced signals considered in Ref.~\cite{Hubert2025.PhysStatSolidiB.262.2400541} cannot tell the spontaneous MOKE response as well as its spatial variation.

Here, in this paper, we report comprehensive MOKE studies on the altermagnet MnTe.
By scanning MOKE, we found two kinds of domains with positive and negative MOKE responses, in full agreement with the TRSB of the altermagnetic order.
We also found that the Kerr angle at low temperature reaches $\pm 10$~mrad, exceeding typical MOKE responses in conventional ferromagnets and exhibiting huge figure of merit toward high-density device assembling.
Theoretical calculation based on first-principles calculations, as well as comparison with MOKE responses in known magnets, demonstrates that the observed MOKE in MnTe is intrinsic to the TRS-broken altermagnetic magnetic structure, with negligible roles of the tiny magnetization originating from canted magnetic components.
The overall results of this study reveal promising potential of MnTe as a material for novel high-density magneto-optical devices.

\section{Methods}\label{sec11}

\subsection{Sample preparation and characterization}

Single crystals of MnTe were grown by the chemical vapor transport method using iodine as the transport agent. 
A 99.9\% purity MnTe lump purchased from Kojundo Chemical Laboratory Co., Ltd. was used as the starting material. 
The growth conditions followed those reported in Ref.~\cite{deMelo1990.JCrystGrowth.104.780}.

We used samples with cleaved $ab$ planes.
The in-plane direction of crystals, as well as crystalline quality, were checked by an x-ray Laue camera (RASCO-BL2, Rigaku).
As shown in Fig.~\ref{fig:sup:Laue} in Appendix, the sample used for the scanning experiment shows clear Laue spots typical for hexagonal systems.
Thus there are no detectable crystalline mosaics. 
Also, we have took Laue photos at various positions on this sample and the quality of the Laue spots is found to be almost the same.
\rev{This fact indicates high homogeneity of the sample.}

Properties of samples grown by the same method are reported in Ref.~\cite{Aoyama2024.PhysRevMaterials.8.L041402}, 
where the N\'eel temperature is reported to be $\TN = 303$~K and the samples show metallic temperature dependence of resistance.

\subsection{Scanning MOKE microscope}

The conventional description of magneto-optical Kerr effect (MOKE) is a rotation of linear polarization upon reflection by an angle $\thetaK$, which is called the Kerr angle.
Similarly, using circularly polarized lights as the bases, we can express MOKE as the phenomena that either of the left- or right-circularly polarized light gain an excessive phase factor $2\thetaK$ upon reflection, as illustrated in Fig.~\ref{fig:fig1}{\bf b}.
These two notations are equivalent for isotropic materials.
However, for materials with orthorhombic anisotropy, the latter is solely sensitive to TRSB, whereas the former can be affected by non-TRSB-related contributions such as birefringence~\cite{Rathgen2005.PhysRevB.72.014451} or unconventional Kerr effect~\cite{li2025quantummetricinducedmagnetooptical,li2025unconventionalmagnetoopticaleffectsaltermagnets}. 
For MnTe, its in-plane N\'eel order lowers the in-plane hexagonal symmetry to orthorhombic, and thus these effects are indeed allowed below $\TN$.
Thus, for detection of TRSB, MOKE measurements using circularly polarized light is essential.
In the present experiment, the magnetic structures enclosed in the red rectangle and those in the blue rectangle in Fig.~\ref{fig:fig1}{\bf b} should exhibit Kerr rotations with different signs.

To measure MOKE, we utilized the all-fiber loop-less Sagnac interferometer.
The original version of the loop-less Sagnac interferometer was described in Refs.~\cite{Xia2006.PhysRevLett.97.167002,Xia2006.ApplPhysLett.89.062508}.
The interferometer was constructed using polarization maintaining (PM) fiber in order to realize two independent optical paths within one fiber.
In our setup, a PM-fiber-based polarizer (PWP-02-15-PM-2-1, Phoenix) and phase modulator (MPX-LN-0.1, iXblue Photonics) were used, instead of the free-space polarizer and modulator.
We use incoherent infrared light from a superluminescent diode (SLD) (S5FC1550S-A2, Thorlabs).
The central wavelength is 1550~nm (corresponding to 0.80~eV photon energy) and the output light power is typically chosen to be 1.5~mW.
The voltage waveform to the phase modulator was applied using a function generator (SDG2082X, Siglent).
The frequency of the phase modulation was chosen to be 1.551~MHz, which was determined from preceding measurements of voltage-frequency characteristics. 
The phase modulation amplitude was set to be 3.127~Vp-p, which realizes the modulation depth of $2\phim = 1.84$. 
The output optical power from the interferometer was measured using a high-resolution InGaAs based photodetector (1811-FC, Newport).
The voltage from the photodetector was analyzed using dual-frequency digital lock-in amplifier (LI5660, NF Corporation).
The readings of the lock-in amplifier is further analyzed using the phase analysis described in the next subsection. 

To achieve high spatial resolution, we used a customized focuser (LPF-D4-1550-8, OZ) with the working distance of 1.8~mm and the focused beam diameter of 2.7~$\mu$m.
\rev{The spatial resolution of our system is evaluated to be $1.16 \pm 0.12~\mu$m (see Appendix).}
A quarter waveplate is fixed beneath the focuser.
The relative angle between the waveplate and focuser is adjusted before scanning measurement so that the optical axis of the waveplate is 45-degree inclined with respect to the optical axes of the PM fiber.

To achieve scanning MOKE measurements, the focuser is placed on our XYZ scanner. 
The XY scan was achieved using an electric XY stage (OSMS20-35(XY)-M6, Sigma Koki) with the minimum step width of 1~$\mu$m and the maximum scan length is 35~mm.
This scanner can be electrically controlled from a computer via a two-axis controller (GSC-02C, Sigma Koki).
The Z scanner (ZSG60, Misumi) only allows control by hand, and is mainly used to optimize the focus of the light.
We always scan along the positive X direction while decreasing the Y position, to avoid possible backlash effects.
After each XY scan, the origin of the XY stage was calibrated.
We typically use scan speed of 1~sec/pixel, while setting the time constant of the lock-in amplifier to be 200~msec.
With these parameters, we have found that measurement delay is negligible. 
Before performing scans of MnTe, the microscope has been tested with ferromagnets Fe and Ni to confirm that the spatial magnetic information can be obtained.
\rev{More details of the MOKE measurements are discussed in Appendix}

Temperature of the sample was controlled by a Peltier temperature controller (UT4070C100F, Ampere Inc.), which allows temperature control between 243~K ($-30^\circ$C) and 353~K  ($+80^\circ$C). 
We placed a sample stage made of oxygen-free copper on the Peltier device. The sample is fixed on a copper plate with silver paint and this plate was fixed onto the sample stage with screws.
In order to apply magnetic field, we used a cylindrical neodymium permanent magnet that can cover the sample stage.
The magnetic field at the sample position was measured to be 0.13~T using a Hall sensor (410 Gaussmeter, LakeShore).

\subsection{Theoretical calculation}

We perform density functional theory (DFT) calculations within the full potential local orbital (FPLO) method~\cite{FPLO}, using the generalized gradient approximation (GGA) under full relativistic condition.
We consider the AFM state shown in Fig.~\ref{fig:fig1}\textbf{a}. This magnetic configuration corresponds to space group $P\bar{3}m1$ (No.~164), 
which is derived from the parent space group $P6_3 /mmc$ (No.~194) of paramagnetic MnTe by removing the glide symmetry connecting two Mn ions along the $c$ axis.
The lattice constants are set to $a=4.14831$\,{\AA} and $c=6.71623$\,{\AA} from Ref.~\cite{Yamamoto2025.PhysRevAppl.24.034037}.
We compute collinear AFM states with two different choices of spin quantization axis, as illustrated in Fig.~\ref{fig:sup:magnetic_structure} \rev{in Appendix}.
\rev{The panel} \textbf{a} corresponds to the experimental \rev{magnetic structure shown} in Fig.~\ref{fig:fig1}\textbf{a}, where the spin moments are oriented along the $[10\bar{1}0]$ direction (the $a^\ast$ direction). On the other hand, \rev{the panel} \textbf{b} shows moments aligned along the $[\bar{1}2\bar{1}0]$ direction (the $a$ direction) for comparison.
The corresponding magnetic point groups are $m'm'm$ and $mmm$ for \textbf{a} and \textbf{b}, respectively.
The spin density is converged self-consistently using a $24 \times 24 \times 24$ $\bm{k}$-point mesh.


Figure~\ref{fig:sup:band} \rev{in Appendix} shows the energy dispersion and the density of states for the $m'm'm$ magnetic structure.
The highest occupied band is mainly composed of Te $5p$ orbitals, whereas the lowest unoccupied band is formed by Mn $3d$ orbitals.
The direct band gap is \rev{0.512}~eV near the A point along $\Gamma$--A line, and the indirect gap is \rev{0.139}~eV between the A and L points.
The magnetic moment converged is \rev{4.45}\,$\mu_\textrm{B}$ per Mn ion.
We construct a tight-binding model using projective Wannier functions within FPLO~\cite{Eschrig2009,Koepernik2023}. The resulting 60-orbital model (including spin degrees of freedom) comprises Mn $3d$, Mn $4s$, Te $5s$, Te $5p$, and Te $5d$ orbitals.
The same procedure is applied to the $mmm$ magnetic structure.


To control the canting moment along the $c$ axis, we introduce an artificial magnetic field acting only on the Mn $3d$ orbitals, described by the Zeeman term
$H_\textrm{Zeeman}=-h \sum_i S_i^z$, where $S_i^z$ denotes the spin operator of the Mn $3d$ orbitals along the $c$ axis.
We obtain a canting moment of $5.4\times10^{-2}\mu_\textrm{B}$ for $h\simeq 432$\,T.
Although the canting moment is exactly zero in the DFT results, the Wannier model exhibits a tiny canting moment of order $10^{-6}\,\mu_\textrm{B}$ even at $h=0$, presumably due to numerical errors associated with the energy-window cutoff.


We calculate the Kerr rotation angle $\theta_\textrm{K}(\omega)$ within the linear response theory using the Kubo formula for the optical conductivity tensor $\sigma_{\alpha\beta}(\omega)$, where $\alpha, \beta=x, y, z$.
We use a $192 \times 192 \times 192$ $\bm{k}$-points for evaluating the eigenenergies of the Wannier tight-binding model.
Once $\sigma_{\alpha\beta}(\omega)$ is obtained, the Kerr rotation angle $\theta_\textrm{K}(\omega)$ is evaluated as
$\theta_\textrm{K}(\omega)=\textrm{Re}\Phi_\textrm{K}(\omega)$,
where the complex Kerr angle is given by~\cite{PhysRev.186.891}
\begin{align}
    \Phi_\textrm{K}(\omega)=-\frac{\sigma_{xy}(\omega)}{\overline{\sigma}(\omega) \sqrt{1+i(4\pi/\omega)\overline{\sigma}(\omega)}}.
    \label{eq:Kerr_calc}
\end{align}
Here, the overline denotes the in-plane average defined as $\overline{\sigma}(\omega) = [\sigma_{xx}(\omega) + \sigma_{yy}(\omega)]/2$.
We have assumed the antisymmetry of the conductivity tensor, $\sigma_{xy}(\omega)=-\sigma_{yx}(\omega)$, in deriving Eq.~\eqref{eq:Kerr_calc}. 
We comment that, for the magnetic configuration in Fig.~\ref{fig:sup:magnetic_structure}\textbf{b}, the presence of the canting moment reduces the symmetry from $mmm$ to $2/m$, which allows a symmetric component, $\sigma_{xy}(\omega)=\sigma_{yx}(\omega)$~\cite{Kleiner1966,Seemann2015}. 
Nevertheless, we neglect the unconventional Kerr contribution~\cite{li2025quantummetricinducedmagnetooptical,li2025unconventionalmagnetoopticaleffectsaltermagnets} arising from this symmetric component.


\section{Results}

\subsection{Magneto-optical imaging of altermagnetic domains in MnTe}

To detect altermagnetism, we employed the polar MOKE, where the wavevectors of incident and reflected lights are both perpendicular to the sample surface.
For this configuration, MOKE is sensitive to the TRSB along the surface normal.
Moreover, for MnTe, MOKE should be expressed as the phenomenon in which either of the left- or right-circularly polarized lights gain an additional phase factor $2\thetaK$ upon reflection depending on the sign of TRSB, as illustrated in Fig.~\ref{fig:fig1}{\bf b}, as discussed in detail in Methods.
Notice that this notation is free from complication originating from non-magnetic contribution such as birefringence~\cite{Rathgen2005.PhysRevB.72.014451} or the symmetric component of the off-diagonal conductivity tensor (so-called unconventional Kerr angle)~\cite{li2025quantummetricinducedmagnetooptical,li2025unconventionalmagnetoopticaleffectsaltermagnets}, which can be non-zero for the in-plane N\'eel order of MnTe.
Using circularly polarized lights, the magnetic structures enclosed in the red rectangle and those in the blue rectangle in Fig.~\ref{fig:fig1}{\bf b} should exhibit Kerr rotations with opposite signs.

In this study, we used a single-crystalline MnTe sample with $\TN =  303$~K~\cite{Aoyama2024.PhysRevMaterials.8.L041402} having high crystalline quality and homogeneity as shown in Fig.~\ref{fig:sup:Laue}. 
The altermagnetic domains are studied using our scanning MOKE microscope.
This microscope utilizes the high-resolution loop-less Sagnac interferometer~\cite{Xia2006.PhysRevLett.97.167002,Xia2006.ApplPhysLett.89.062508}.
In this interferometer, just before the light is shined to the sample, the light is converted to circularly polarized lights and the interferometer detects sensitively the phase difference between the left- and right-circularly polarized lights, solely detecting TRSB.
The wavelength of our optics centers at 1550~nm (corresponding to the photon energy of 0.80~eV).
The light-spot radius and the spacial resolution of the scanner is 2~$\mu$m as examined in Fig.~\ref{fig:sup:spotsize}\textbf{d}.
Details of this MOKE microscope are explained in Methods.

\begin{figure} 
	\centering
	\includegraphics[width=1\textwidth]{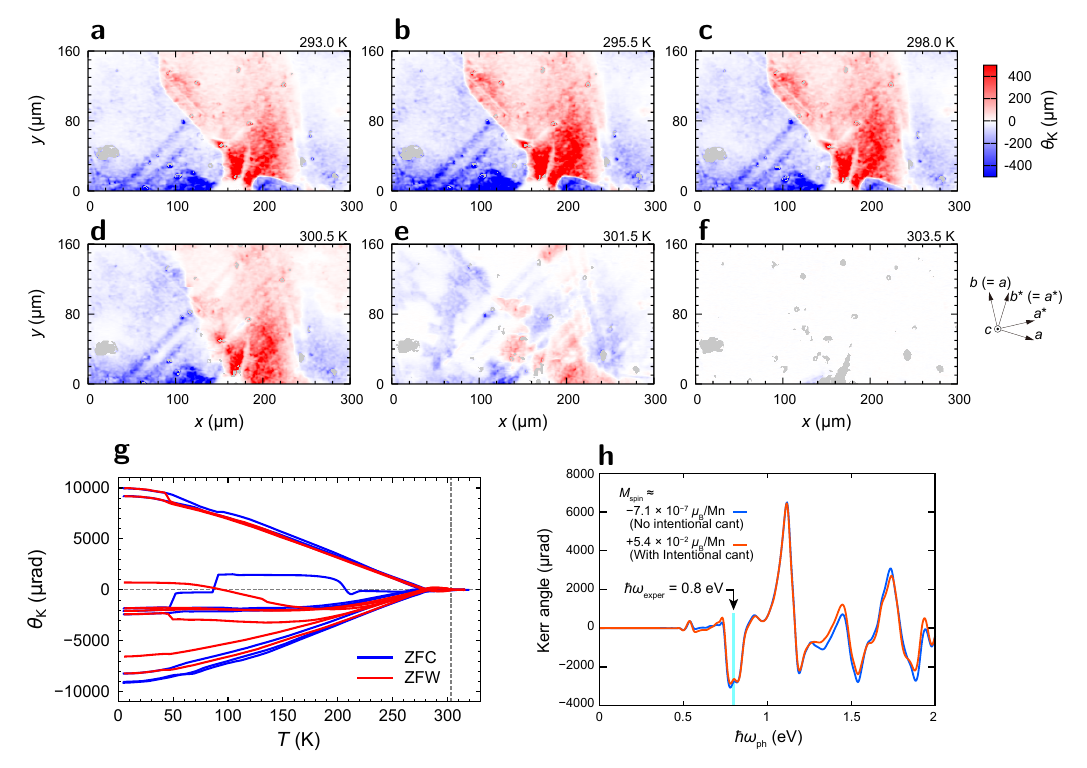} 

	\caption{\textbf{Spontaneous Kerr rotation due to the altermagnetic order of MnTe.}
	\textbf{a}-\textbf{f}. 1-$\mu$m-step Kerr mappings measured at various temperatures shown in the panels. 
    The pixels with $V_{2f} < 15$~mV are shown with gray. 
    After cooling to 293~K, each scan was performed while warming.
    The domain structure weakens upon warming and completely disappears at $\TN = 303$~K.
    \textbf{g}. Temperature dependence of the spontaneous Kerr signal observed in single-spot measurements.
    We here show data of multiple subsequent thermal cycles.
    Zero-field cooling processes (ZFC; blue curves) were followed by zero-field warming processes (ZFW; warming). 
    The Kerr rotation exhibit finite value below $\TN$ (the dashed line), but their sign and magnitude are rather random.
    The low-temperature value of $|\thetaK|$ reaches $\pm 10000$~$\mu$rad.
	\textbf{h}. Theoretical photon-energy $\hbar\omega\subm{ph}$ dependence of the Kerr angle evaluated based on first-principles band calculations.
    Calculation without (blue curve) and with (orange curve) an intentional magnetic-moment canting shows little difference, and both shows finite Kerr angle at various $\hbar\omega\subm{ph}$, including at the experimental photon energy of $\hbar\omega\subm{exper}=0.8$~eV indicated by the cyan line.
    This comparison indicates that the observed Kerr angle is intrinsic to the altermagnetism, but not attributable to the canted magnetization.
    }
	\label{fig:fig2} 
\end{figure}

\begin{figure}[h] 
	\centering
    \includegraphics[width=1.0\textwidth]{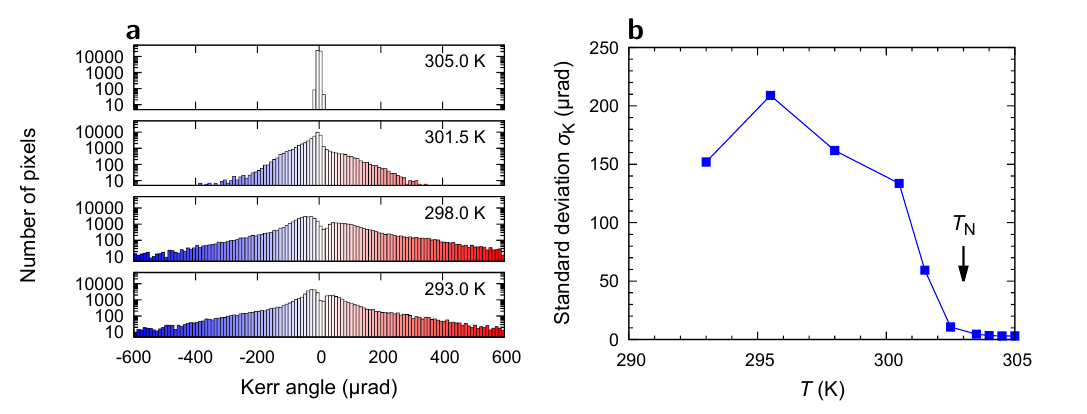} 
	\caption{\textbf{Temperature evolution of the real-space Kerr-angle distribution.}
    \textbf{a}. Histogram of the number of pixels in scans represented in Figs.~\ref{fig:fig2}\textbf{a}-\textbf{f}. 
    The histogram bin is chosen to be 10.0~$\mu$rad and pixels with sufficiently large reflection ($V_{2f}>15$~mV) are only counted.
    \textbf{b}. Temperature dependence of the standard deviation $\sigma\subm{K}$ of the distribution shown in \textbf{a}.
    This quantity roughly correspond to the half width of the distribution peak.
    Notice that $\sigma\subm{K}$ almost disappears at $\TN$.
	}
	\label{fig:sup:histogram}
\end{figure}

Figures~\ref{fig:fig1}\textbf{c} and \textbf{d} show reflectivity and Kerr mappings (with 6-$\mu$m steps) of our MnTe sample taken at 293~K (20$^\circ$C), which is below $\TN$.
Within the sample, we clearly see distinct regions with positive and negative Kerr angles, indicative of formation of altermagnetic domains.
Some of the domains have sizes close to 1~mm.
With increasing temperature, as shown in 1-$\mathrm{\mu m}$-step mappings in Figs.~\ref{fig:fig2}{\bf a}-{\bf f}, the MOKE response becomes weaker and completely vanishes above 303~K, which equals $\TN$ of similar  samples~\cite{Aoyama2024.PhysRevMaterials.8.L041402}.
The disappearance of altermagnetic order is quantitatively investigated using the temperature dependence of the spatial distribution of $\thetaK$ shown in Fig.~\ref{fig:sup:histogram}.
These results show that the observed domains originate intrinsically from the altermagnetic order of MnTe, not from extrinsic origins such as surface contamination.

\subsection{Gigantic Kerr effect at low temperatures}

To extend the temperature range of our experiment, we performed  single-point MOKE measurements using a commercial cryostat (PPMS, Quantum Design). 
For this measurement, we used a compact sample fixture based on ceramic ferrules~\cite{Ikeda2026.PhysRevRes.8.013169}.
The diameter of the focused light is 6~$\mu$m.
We used another single crystal from the same batch as the crystal used for scanning.
The sample is placed inside a permalloy magnetic shield in order to minimize the remnant field around the sample~\cite{Yamane2026.arXiv2606.10300}.

We repeated several temperature sweeps between 5~K and 320~K at zero field.
Representative results are shown in Fig.~\ref{fig:fig2}\textbf{g}.
Below $\TN = 303$~K, the Kerr angle changes substantially but the change has random nature: in some sweeps $\thetaK$ exhibits positive changes, in some sweeps negative changes, and in other sweeps $\thetaK$ exhibits sudden sign change during the sweep.
This random nature provides strong additional evidence for spontaneous nature of the altermagnetic TRSB in MnTe.

\begin{figure} 
	\centering
	\includegraphics[width=0.5\textwidth]{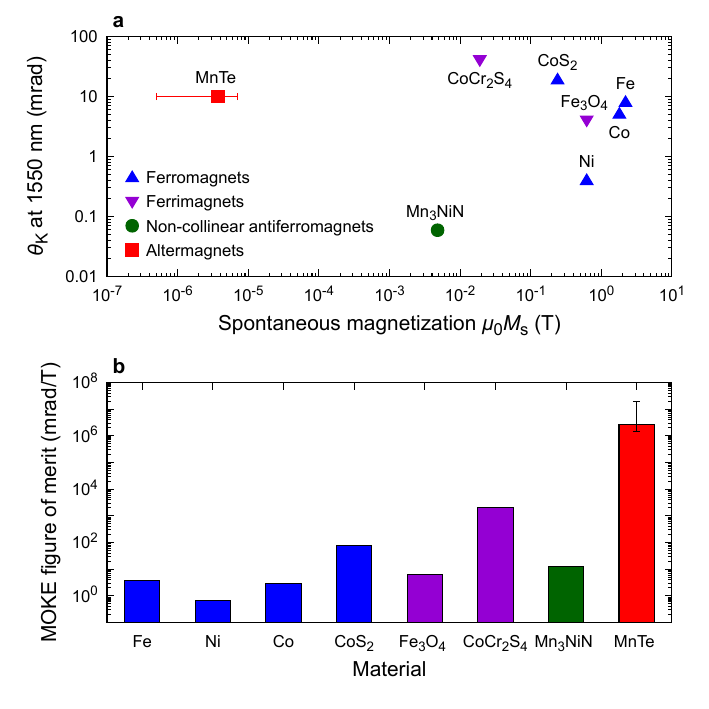} 
	\caption{\textbf{Extremely high MOKE performance of MnTe.}
 	\textbf{a}. Relation between MOKE measured at 1550-nm wavelength and spontaneous magnetization $M\subm{s}$ in various magnetic materials~\cite{Buschow1983.JMagMagMat.38.1, Delin1999.PhysRevB.60.14105, Sato1982.JPhysSocJpn.51.2955, Miyahara1968.JApplPhys.39.896,Fontijn1997.PhysRevB.56.5432, Ohgushi2008.JPhysSocJpn.77.034713, Farhang2025.arXiv.2510.19709, Wu2013.JApplPhys.114.123902, Aoyama2024.PhysRevMaterials.8.L041402, Kluczyk2024.PhysRevB.110.155201}. 
    Numerical values and their references are given in Table~\ref{tab:M-Kerr}.
    Ferromagnets (blue up triangles) and ferrimagnets (purple down triangles) exhibits large MOKE of around 1--10~mrad but with large spontaneous magnetization of the order of 1~T.
    Non-collinear antiferromagnets (green circles) exhibits lower MOKE and magnetization. 
    In contrast, the altermagnet MnTe (red square) is found to exhibit MOKE comparable to those in ferromagnets or ferrimagnets, whereas $M\subm{s}$ is million times smaller.
    \textbf{b}. MOKE figure of merit (FOM), defined as $|\thetaK|/(\mu_0 M\subm{s})$, of various magnets. 
    This value characterize the possibility of high-density device assembly without affected by stray fields while maintaining high MOKE responses.
    MnTe is found to exhibit exceptionally high FOM, demonstrating applicability to high-density and high-performance MO devices.
       }
	\label{fig:fig3} 
\end{figure}

Moreover, we found that the value of $\thetaK$ reaches $\pm 10$~mrad in some sweeps at lowest temperatures (Fig.~\ref{fig:fig2}\textbf{g}). 
In MnTe, it is known that spin-orbit coupling cants the magnetic moments, resulting in a tiny  spontaneous magnetization of the order of $10^{-5}$ -- $10^{-6}~\mu\subm{B}$/Mn along the $c$ axis~\cite{Kluczyk2024.PhysRevB.110.155201,Aoyama2024.PhysRevMaterials.8.L041402}.
In spite of such tiny spontaneous magnetization, the observed $\thetaK$ values are comparable to those of ferro- or ferri-magnets, as shown in Fig.~\ref{fig:fig3}\textbf{a} and Table~\ref{tab:M-Kerr}.

We then compare $\thetaK$ and spontaneous magnetization of various magnets quantitatively.
As shown in Fig.~\ref{fig:fig3}\textbf{a}, standard ferromagnets and ferrimagnets exhibits saturated magnetization per unit volume ($\mu_0M$) of the order of 1~T and the Kerr rotation at 1550~nm of the order of 1--10~mrad~\cite{Buschow1983.JMagMagMat.38.1,Delin1999.PhysRevB.60.14105,Fontijn1997.PhysRevB.56.5432}.
The half-metal candidate CoS\sub{2} and the ferrimagnet CrCo\sub{2}S\sub{4} exhibit large MOKE exceeding 10~mrad due to the resonance to the d-d transition, in spite of relatively smaller magnetization~\cite{Sato1982.JPhysSocJpn.51.2955,Miyahara1968.JApplPhys.39.896,Ohgushi2008.JPhysSocJpn.77.034713}. 

\begin{table}[h]
	\centering
	\caption{\textbf{Comparison of spontaneous Kerr angle and magnetization in various magnetic materials breaking time-reversal symmetry.}
    In this table, we list only the data taken with infrared light of 1550-nm wavelength (corresponding to 0.80-eV photon energy). 
	The spontaneous magnetization per unit volume, $M\subm{s}$, is expressed in the unit of tesla by multiplying $M\subm{s}$ by $\mu_0$. 
    FM refers to ferromagnet and AFM refers to antiferromagnet.
    MnTe exhibits a gigantic MOKE figure of merit (FOM), which is defined by $|\thetaK|/(\mu_0 M\subm{s})$.}
	\label{tab:M-Kerr} 
    {\small 
	\begin{tabular}{ccccc} 
		\\

   \hline
		Material & Type & $\thetaK$ (mrad) & $\mu_0M\subm{s}$ (T) & FOM (mrad/T) \\
		\hline
		Fe                  & FM & $-8.0$ \cite{Buschow1983.JMagMagMat.38.1, Delin1999.PhysRevB.60.14105} & 2.2 & $3.6$ \\
		Ni                  & FM & $+0.4$ \cite{Buschow1983.JMagMagMat.38.1, Delin1999.PhysRevB.60.14105}& 0.62 & $6.4$ \\
		Co                  & FM & $-5.1$ \cite{Buschow1983.JMagMagMat.38.1, Delin1999.PhysRevB.60.14105}& 1.8 & $2.8$ \\
	    CoS\sub{2}          & FM & $+19 $ \cite{Sato1982.JPhysSocJpn.51.2955}& 0.24~\cite{Miyahara1968.JApplPhys.39.896} & $78$ \\
		Fe\sub{3}O\sub{4} & ferrimagnet & $-4.0$~\cite{Fontijn1997.PhysRevB.56.5432} & $0.62$ & $6.5$ \\
		CoCr\sub{2}S\sub{4} & ferrimagnet & $-41$~\cite{Ohgushi2008.JPhysSocJpn.77.034713} & $0.019$ & $2100$ \\
        Mn\sub{3}NiN & Non-collinear AFM & $+0.059$~\cite{Farhang2025.arXiv.2510.19709} & $4.8\times 10^{-3}$ ~\cite{Wu2013.JApplPhys.114.123902} & $12$ \\
        MnTe & altermagnet & $- 10$ (This work) & $5\times 10^{-7}$~\cite{Aoyama2024.PhysRevMaterials.8.L041402} &  $2.0\times 10^7$ \\
        MnTe & altermagnet & $- 10$ (This work) & $7\times 10^{-6}$~\cite{Kluczyk2024.PhysRevB.110.155201} & $1.4\times 10^{6}$ \\
		\hline
        
	\end{tabular}
    }
\end{table}

The magnetization is a source of stray fields, which is the major constraint toward high-density device assembly.
To realize MOKE-based devices such as MO memories, large MOKE is advantageous for reliable operations.
Thus, the ratio $|\thetaK|/(\mu_0 M\subm{s})$ can be used as a figure of merit (FOM) towards realization of high-density and high-performance MO devices.
Figure~\ref{fig:fig3}\textbf{b} shows this FOM. 
Ordinary ferromagnets and ferrimagnets exhibits FOM of the order of 1~mrad/T, whereas the FOM of CoS\sub{2} and CrCo\sub{2}S\sub{4} amounts $10^2$ -- $10^3$~mrad/T.
In antiferromagnets breaking TRS, such as the non-collinear antiferromagnet Mn\sub{3}NiN, their properties (59~$\mu$rad at 1550~nm, $\mu_0 M \sim 0.0048$~T)~\cite{Wu2013.JApplPhys.114.123902, Farhang2025.arXiv.2510.19709} results in FOM of 12~mrad/T, which is still comparable to those of ferroic magnets. 
For MnTe, the magnetization due to the canted moment in the unit of tesla is as small as $0.5\times 10^{-6}$ -- $7.0 \times 10^{-6}$~T\cite{Kluczyk2024.PhysRevB.110.155201,Aoyama2024.PhysRevMaterials.8.L041402}.
Thus, $\thetaK$ of $\pm 10$~mrad at low temperature leads to huge FOM ranging in $1.4\times 10^{6}$ -- $2.0 \times 10^{7}$~mrad/T, demonstrating the potential of MnTe as a material for compact MO devices.

\subsection{Origin of the large Kerr rotation in MnTe}

The extremely high FOM indicates unconventional mechanism of MOKE in MnTe, rather than the conventional MOKE originating from magnetization.
Indeed, as described below, our theoretical calculation reveals that the gigantic MOKE in MnTe originate intrinsically from the altermagnetic order.
The band structure and corresponding optical processes are evaluated based on the band structure obtained by density-functional theory (DFT) calculation (Fig.~\ref{fig:sup:band} in Appendix) and the Kerr angle is evaluated assuming various magnetic structures.
Details of the calculation are described in Methods.

\begin{figure} 
	\centering
    \includegraphics[width=\textwidth]{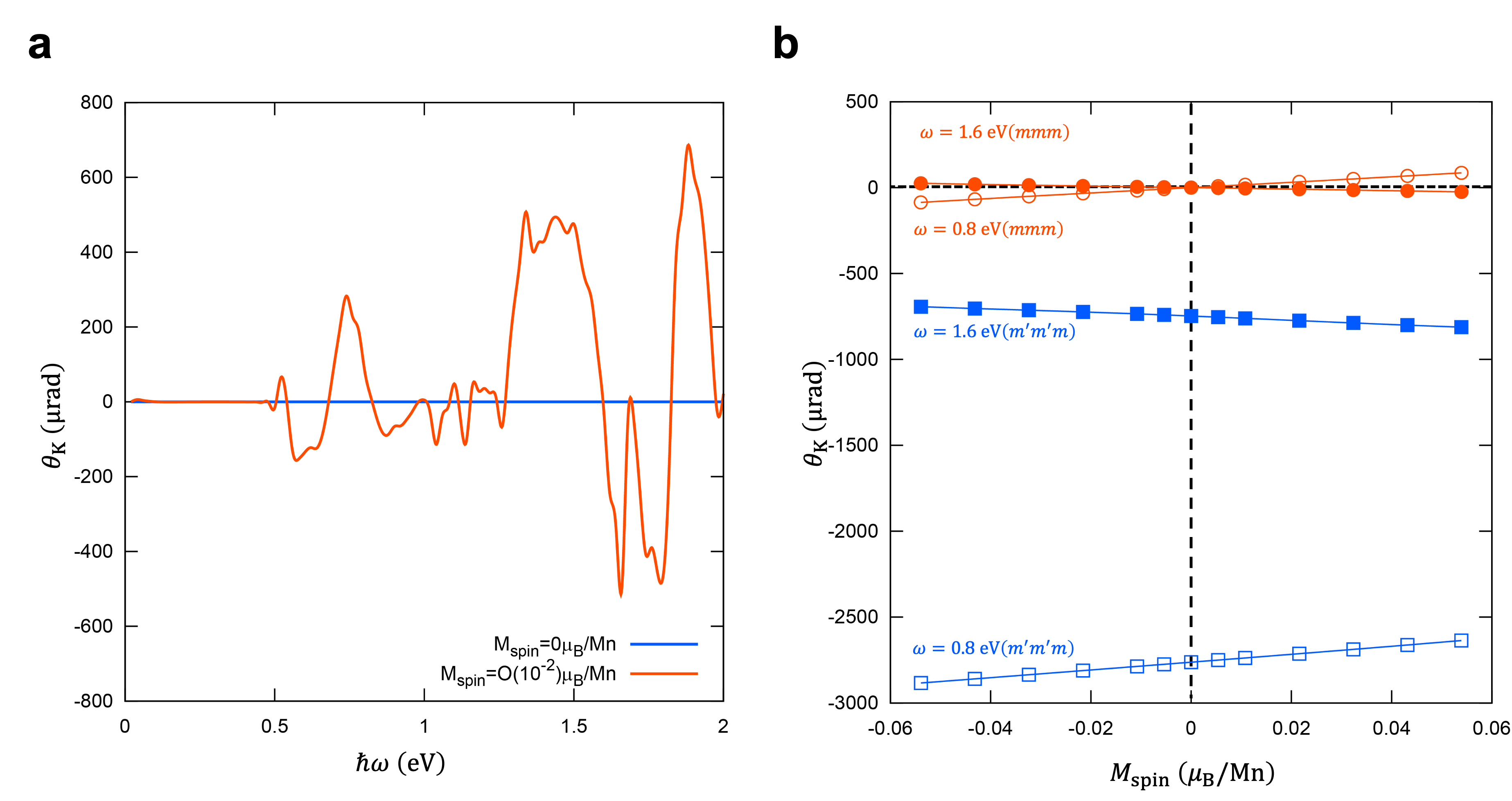}
	\caption{\textbf{Calculated Kerr angle $\theta_\textrm{K}(\omega)$ for $m'm'm$ and $mmm$ magnetic structures.}
    (\textbf{a}) Photon-energy $\hbar\omega$ dependence of $\theta_\textrm{K}(\omega)$ for $mmm$ structure. The result for $m'm'm$ is presented in Fig.~\ref{fig:fig2}\textbf{g}.
    (\textbf{b}) Canting-moment $M_\textrm{spin}$ dependence of $\theta_\textrm{K}(\omega)$ for fixed photon energy. The blue squares and red circles show the results for the $m'm'm$ and $mmm$ magnetic structures, respectively.
    Here, we assume that the N\'eel-order structure does not flip under negative $M\subm{spin}$.
	}
	\label{fig:sup:Kerr_angle_mmm}
\end{figure}

When magnetic moments align along the $a^\ast$ direction (the $m'm'm$ structure; Fig.~\ref{fig:sup:magnetic_structure}{\bf a} in Appendix), MnTe exhibits finite Kerr angles at various photon energy $\hbar\omega\subm{ph}$ even for the collinear N\'eel order obtained directly from the DFT calculation (the blue curve in Fig.~\ref{fig:fig2}{\bf h}).
At the experimental photon energy of 0.8~eV, $\thetaK$ is calculated to be $-2.8$~mrad, which is comparable to the observed low-temperature value ($\pm 10$~mrad).
We comment that this structure contains tiny net magnetization of the order of $10^{-7}~\mu\subm{B}$/Mn due to a limitation in calculation resolution.
To examine the role of canted magnetization, we also performed calculations with intentional magnetization canting.
The result with a net magnetization of $5.4\times 10^{-2}~\mu\subm{B}$/Mn (orange curve in Fig.~\ref{fig:fig2}{\bf g}) shows little difference from the blue curve, in spite of difference in magnetization by 5 orders of magnitude.

\rev{
We have also checked that the Kerr angle is zero if magnetic moments align along the $a$ axis (the $mmm$ structure; see Fig.~\ref{fig:sup:magnetic_structure}{\bf b} in Appendix).
Figure~\ref{fig:sup:Kerr_angle_mmm}\textbf{a} shows the Kerr angle $\theta_\textrm{K}(\omega)$ for the $mmm$ magnetic structure. 
In contrast to the $m'm'm$ case (Fig.~\ref{fig:fig2}\textbf{h}), the Kerr angle for $mmm$ is zero (blue curve), and exhibits finite values only when the canting moment along the $c$ axis is artificially introduced (red curve).
This behavior can be understood in symmetry point of view. 
The $mmm$ symmetry does not allow anomalous Hall conductivity
~\cite{Kleiner1966.PhysRev.142.318}, thus nor spontaneous MOKE. 
However, the canting moment lowers the symmetry from $mmm$ to $m'm'm$ or even lower, thereby allowing a nonzero optical Hall conductivity $\sigma_{xy}(\omega)\neq 0$, and consequently, a nonzero Kerr rotation.
To further highlight the difference between the magnetic structures, the canting-moment $M_\textrm{spin}$ dependence of the Kerr angle is shown in Fig.~\ref{fig:sup:Kerr_angle_mmm}\textbf{b}.
Here, although negative $M\subm{spin}$ would results in the change in the N\'eel-order structure to its time-reversal counterpart in actual samples (see Fig.~\ref{fig:fig1}{\bf b}), such an effect is not considered in this calculation.
It is clearly seen that $\theta_\textrm{K}(\omega)$ remains finite even in the absence of $M_\textrm{spin}$ for the $m'm'm$ structure, whereas it vanishes as $M_\textrm{spin} \to 0$ for the $mmm$ structure.
}

\begin{figure} 
	\centering
    \includegraphics[width=\textwidth]{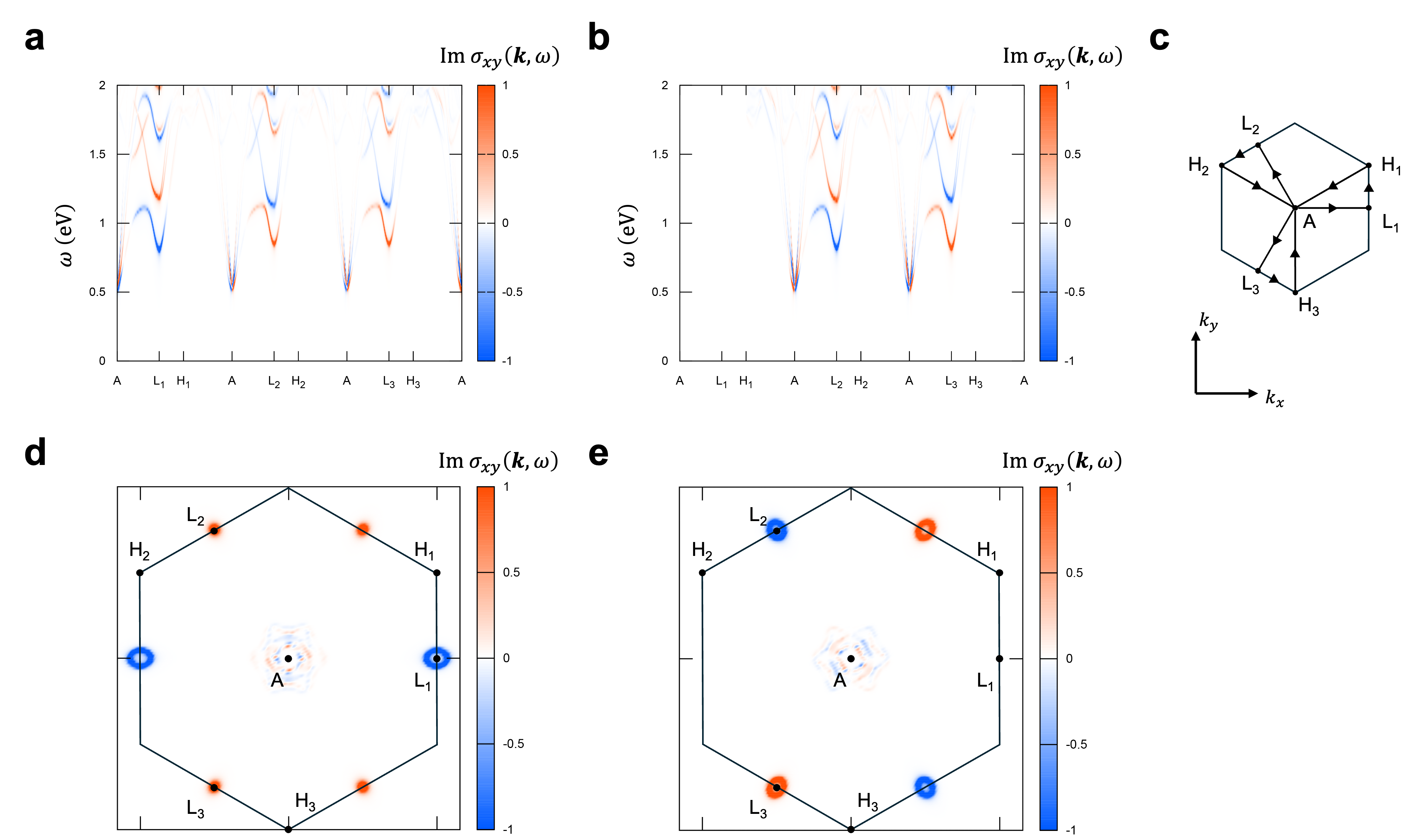}
	\caption{\textbf{Momentum-resolved optical Hall conductivity $\sigma_{xy}(\bm{k},\omega)$ for $m'm'm$ and $mmm$ structures}
    (\textbf{a},\textbf{b}) Momentum-resolved $\mathrm{Im}\sigma_{xy}(\bm{k},\omega)$ for $m'm'm$ and $mmm$ in arbitrary units, respectively, along with three equivalent paths in \textbf{c}.
    (\textbf{c}) Path in the Brillouin zone.
    (\textbf{d},\textbf{e}) $\mathrm{Im}\sigma_{xy}(\bm{k},\omega)$ on a $k_z=\pi/c$ plane at $\hbar\omega=0.85$\,eV for $m'm'm$ and $mmm$, respectively.
	}
	\label{fig:sup:optical_conductivity_BZ}
\end{figure}

Finally, we discuss the anisotropy of the electronic structure that gives rise to the finite Hall conductivity. 
To this end, we introduce the momentum-resolved optical Hall conductivity $\sigma_{xy}(\bm{k},\omega)$, which is related to the total Hall conductivity via the sum over all momenta $\bm{k}$ of electrons as
$\sigma_{xy}(\omega) = \sum_{\bm{k}} \sigma_{xy}(\bm{k},\omega).$
Figures~\ref{fig:sup:optical_conductivity_BZ}\textbf{a} and \ref{fig:sup:optical_conductivity_BZ}\textbf{b} show $\sigma_{xy}(\bm{k},\omega)$ along three symmetry-equivalent paths defined in Fig.~\ref{fig:sup:optical_conductivity_BZ}\textbf{c}.
The three equivalent L points have contributions at $\hbar\omega \approx 0.85$\,eV with different sign.
Figures.~\ref{fig:sup:optical_conductivity_BZ}\textbf{d} and \ref{fig:sup:optical_conductivity_BZ}\textbf{e} show the $\bm{k}$ dependence at $\hbar\omega=0.85$\,eV.
In Fig.~\ref{fig:sup:optical_conductivity_BZ}\textbf{d}, the three symmetry-equivalent L points, denoted $\mathrm{L}_1$, $\mathrm{L}_2$, and $\mathrm{L}_3$, yield negative, positive, and positive contributions, respectively. As a result, their sum yields a finite net value of $\sigma_{xy}(\omega)$ for the $m'm'm$ structure.
In contrast, in Fig.~\ref{fig:sup:optical_conductivity_BZ}\textbf{e}, $\mathrm{L}_1$ does not contribute, while $\mathrm{L}_2$ and $\mathrm{L}_3$ produce contributions of opposite sign that cancel each other. Consequently, the total Hall conductivity $\sigma_{xy}(\omega)$ vanishes for the $mmm$ structure.

\rev{
These examinations indicate that the TRS-broken altermagnetic order is the origin of the observed Kerr rotation, and the canted moment plays negligible roles.
The spontaneous Kerr angle originates mainly from optical transitions at the $L$ points in the Brillouin zone, with avoided cancellation of $k$-resolved optical Hall conductivity due to the broken mirror symmetry with respect to the $k_y = 0$ plane.
}

\subsection{Thermal and magnetic control of altermagnetic domains}

Domain stability against thermal perturbations provides fundamental information toward application of altermagnetism.
Figures~\ref{fig:fig2}\textbf{a}--\textbf{f} show that the domain structure is almost unchanged up to 298~K, manifesting stability of devices up to this temperature. 
However, above 300~K, domain configuration changes in several parts of the sample: e.g. the blue part at around $(x, y) = (200~\mathrm{\mu m}, 0~\mathrm{\mu m})$ in \textbf{a}--\textbf{c}\ changes into red in \textbf{d}, indicating that thermal energies overcome domain-wall pinning energies in some parts of the sample.

\begin{figure} 
	\centering
	\includegraphics[width=0.8\textwidth]{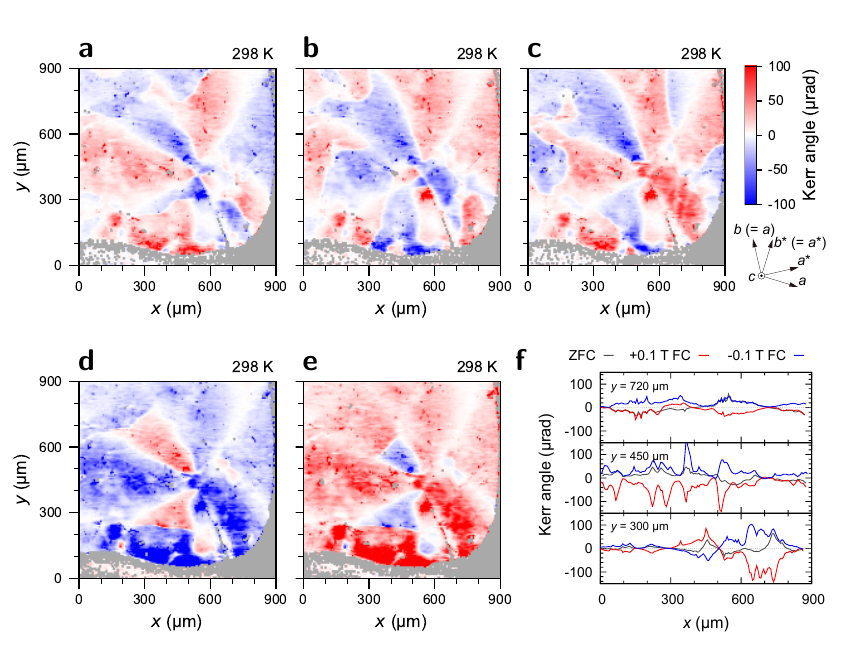} 

	\caption{\textbf{Thermal and magnetic control of the altermagnetic domain in MnTe.}
    \textbf{a}-\textbf{c}. 6-$\mu$m-step Kerr mappings taken after warmings above $\TN$. 
    Pixels with $V_{2f} < 30$~mV are shown with gray. 
    Parts of the domain configuration changes by thermal cycles, indicating spontaneous nature of the altermagnetic order.
    \textbf{d}. Kerr mappings taken after cooling in positive magnetic fields of around $+0.1$~T. 
    \textbf{e}. Same mapping upon $-0.1$-T field cooling.
    Majority of the domains are trained by field coolings. 
    \textbf{f}. Line cuts of the images \textbf{a} (gray curves), \textbf{d} (red curves), and \textbf{f} (blue curves).
    The $x$ dependences of the Kerr angle at various $y$ positions indicated in the panels are shown.
    In many parts of the sample, the magnetic trainings alter the sign of the Kerr angle while preserving detailed structures in the magnitude of the Kerr angle.
    }
	\label{fig:fig4} 
\end{figure}

To further uncover thermal effects on the domain formation, we repeated MOKE scannings after we warmed the sample up to above $\TN$ and cooled it to 298~K.
As shown in Figs.~\ref{fig:fig4}{\bf a}-{\bf c}, domain patterns are largely altered once the sample is warmed up above $\TN$.
For example, the blue triangle-like domain in the right of the panel {\bf a} turns into red in {\bf b}, and then changes to a mixed configuration of the red and blue domains in {\bf c}.
This ``random'' hysteretic domain formation provides firm evidence for spontaneous nature of the TRSB.
On the other hand, some features, such as the domain wall (white region) running from the center to the right top corner, are preserved even after repeated thermal cycles.
This fact implies that there are certain domain-wall pinning mechanisms, such as crystal imperfections or local strains, which fixes some of the domain walls even after thermal cycles. 

Magnetic fields allow deterministic domain control, which is  essential for device application.
To demonstrate that, we performed MOKE scanning after cooling the sample across $\TN$ under $c$-axis magnetic fields.
As shown in Figs.~\ref{fig:fig4}{\bf d} and {\bf e}, domain configuration largely altered by field cooling: most part of the sample shows negative Kerr rotation after +0.1~T cooling, whereas it shows positive Kerr rotation after $-0.1$~T cooling. 
These results clearly indicate that the altermagnetism in MnTe couples with external $c$-axis magnetic fields.
Although this resembles ferromagnets, one should recall that the N\'eel order in MnTe consists of in-plane magnetic moments, as illustrated in Figs.~\ref{fig:fig1}{\bf a} and {\bf b}.

Figure~\ref{fig:fig4}{\bf f} shows horizontal line-cuts of Kerr mappings at several $y$ positions for the field-control experiments (Figs.~\ref{fig:fig4}{\bf a}, {\bf d}, and {\bf e}).
Within a domain, there exists several fine structures.
Interestingly, most of the fine structures are preserved after field cooling: many negative peaks in red curves ($+0.1$~T field cooling) changes to positive peaks in blue curves ($-0.1$~T field cooling).
In addition, many of the domain walls, namely the nodes in these curves, are common under positive and negative field trainings.
These facts provide strong evidence that the magnitude of the Kerr response, as well as the position of the domain walls, are determined by local sample properties (such as local defects or strain), but the sign of the Kerr response is easily altered by $c$-axis magnetic field.
This domain-wall stability by local properties while maintaining high TRSB controllability are crucial hints toward building reliable devices.

\section{Discussion and outlook}

Summarizing our findings, we observed mm-sized domains formed due to the spontaneous altermagnetic order in MnTe by high-resolution scanning MOKE measurements.
The Kerr angle was found to reach as large as 10~mrad at low temperature.
Theoretical calculation, as well as comparison with typical magnetic materials, indicates that the observed Kerr rotation is solely due to TRSB of altermagnetism. 
The gigantic MOKE FOM, $|\thetaK|/(\mu_0 M)$, manifests that the altermagnetic structure in MnTe is suitable for high-density MO devices without unwanted stray-field interferences.
We find that the domain configuration can be controlled either by thermal energies or by external vertical magnetic fields.

We should comment on the recent XMCD/XMLD work using thin film (30~nm thickness), showing altermagnetic domains of the order of 1~$\mu$m~\cite{Amin2024.Nature.636.348}. 
Domains observed in our work are much larger.
On the other hand, we also observed fine sub-domain structures (Appendix~\ref{sec:domain-wall} and Figs.~\ref{fig:sup:spotsize}{\bf a}--{\bf c}).
These facts can be explained if we consider a depth-dependent domain structures. 
We propose that altermagnetic domains tend to form fine structures of 1-$\mu$m scale near the surface of a sample, whereas the domains become 1-mm scale in the bulk.
Then, in thin films, only the fine structures can be seen. 
In bulk samples, a combination of the fine and course structures should be observed. 
Our MOKE study uses 0.8~eV infrared light, which typically penetrates into a metal or semiconductor by 10 -- 500~nm~\cite{OpticsHandbook}.
Our calculation reveals that the imaginary part of the complex refractive index, $\kappa$, is 1.31 at the wavelength $\lambda = 1550$~nm, yielding the penetration depth $\lambda/(4\pi \kappa)$ of 94~nm.
Thus, if small domains are formed near the surface whereas larger domains are formed several tenth of nm beneath the surface, then the observed complicated structures consisting of large domains and sub-domain structures can be naturally explained.

Our demonstration of readability of TRSB information by an inexpensive and safe in-house optics using telecom infrared wavelength opens a route toward direct usage of altermagnetism for magneto-optical devices such as optical memories.
The domain controls by thermal and magnetic stimuli provides fundamental bases toward full utilization of altermagnetism, as well as toward understanding of the altermagnetic order itself.
These controllability and stability of altermagnetic domains becomes important hints toward building reliable, tunable, and high-density altermagnet devices.

\bibliographystyle{./apsrev4-2_no-lowercase.bst}

\bibliography{altermagnetism.bib,Kerr.bib,theory.bib}

\clearpage

\begin{acknowledgments}
We acknowledge I.~Kakeya and Y.~Gotoh for valuable comments and discussions.
We also acknowledge Y.~Maeno for providing the Peltier temperature controller.
We also thank Y. Hu, K.~Yada, J.~Xia, C.~Farhang, J.~Wang for technical assistance in the early stage of the introduction of MOKE technique to the Kyoto Univ. Group.
We acknowledge support for the construction of experimental setups from Research Equipment Development Support Room of the Graduate School of Science, Kyoto University; and support for liquid helium and nitrogen supplies from Low Temperature and Materials Sciences Division, Agency for Health, Safety and Environment, Kyoto University.
Part of the computations in this work were performed using the facilities of the Supercomputer Center, the Institute for Solid State Physics, the University of Tokyo (Grants No.~2025-Ca-0086, No.~2025-Cb-0049).

This work was supported by Grant-in-Aids for Academic Transformation Area Research (A) ``Quantum Asymmetry’’ (KAKENHI Grant Nos.~JP23H04866, JP23H04869, JP24H01663) from the Japan Society for the Promotion of Science (JSPS), 
by Grant-in-Aids for Scientific Research (KAKENHI Grant Nos.~JP23K17670, JP23H04861, JP24H00194, JP25K00960, JP25K00955, JP25H01246, JP25H01247) from JSPS,
by ISHIZUE 2023 of Kyoto University Research Development Program, 
by Iketani Science and Technology Foundation (Grant No.~0361078-A),
and by The Mitsubishi Foundation (Grant No.~202410051).
\end{acknowledgments}

\clearpage


\appendix

\section{Sample characterization}

\begin{figure}[h] 
	\centering
	\includegraphics[width=1.0\textwidth]{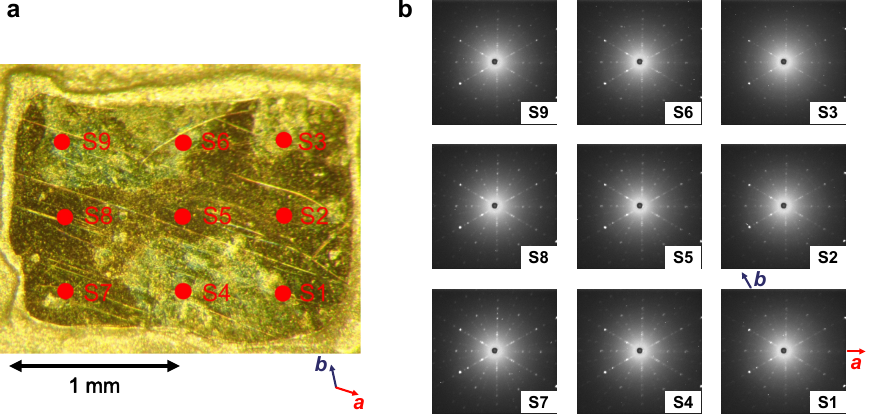} 
	\caption{\textbf{Laue photos of the MnTe sample}
        (\textbf{a})
		Optical image of the single-crystalline MnTe sample used for the present study. The circles indicate positions where Laue photos were taken.
        Directions of the crystalline $a$ and $b$ axes are shown by the arrows.
        (\textbf{b})
        Laue photos of the sample taken at positions indicated in the right-bottom corner of each photos. 
        At all positions, the Laue photos exhibits six-fold pattern consisting of clear spots, indicating high crystalline quality and homogeneity. 
        A comparison with simulation determines the $a$ and $b$ axes are as indicated by the arrows.
        Notice that, when we mounted the sample to the Laue camera, the sample is rotated by around 17$^\circ$ with respect to the photo {\bf a}.}
	\label{fig:sup:Laue} 
\end{figure}

\rev{By using an x-ray Laue camera (RASCO-BL2, Rigaku), we checked the in-plane direction of MnTe crystals, as well as crystalline quality.
As shown in Fig.~\ref{fig:sup:Laue}, the sample used for the scanning experiment shows clear Laue spots with little variations, indicating high crystalline quality and homogeneity.}

\section{Details of the MOKE microscope}

\subsection{Phase analysis}

For MOKE mappings of spontaneous magnetic domains, the Kerr angle can take both positive and negative values depending on local magnetic structures. 
Thus, the sign of the Kerr angle needs to be obtained, in addition to its amplitude.

To resolve the sign of the obtained Kerr signal, the phase analysis of the photo voltage is required. 
More general discussion on this analysis is given in Ref.~\cite{Ikeda2026.PhysRevRes.8.013169}.
The lock-in amplifier provides $V_{1,x}$, $V_{1,y}$, $V_{2,x}$, and $V_{2,y}$, where $V_{n,x}$ ($n = 1,2$) is the magnitude of the $\sin(2\pi nft)$ component of the oscillatory photodetector voltage and  $V_{n,y}$ is that of the $\cos(2\pi nft)$ component.
Considering a phase shift $\delta_n$ due to an arbitrary choice of the origin of the time $t_0$, these voltages have relations to the total magneto-optical (MO) rotation angle $\thetaT$ (Kerr angle plus background) as
\begin{align*}
    V_{x,1} &=2V\subm{m} J_1(2\phim)\sin(2\theta\subm{tot}) \cos\delta_1, \\ 
    V_{y,1} &=2V\subm{m} J_1(2\phim)\sin(2\theta\subm{tot})\sin\delta_1, \\
    V_{x,2} &=2V\subm{m} J_2(2\phim)\cos(2\theta\subm{tot}) \cos\delta_2, \\ 
    V_{y,2} &=2V\subm{m} J_2(2\phim)\cos(2\theta\subm{tot})\sin\delta_2,
\end{align*}
where $V\subm{m}$ is the coefficient to define voltage oscillation amplitude, $J_n(x)$ is the $n$-th order Bessel function of the first kind, and $\phim$ is the depth of the phase modulation.
For MOKE measurements, $\thetaT$ is always much smaller than 1 and thus the 2nd harmonic signal is much stronger than the 1st harmonic signal.
Then, the phase shift $\delta_2$ can be accurately determined from the 2nd harmonic signal as
\begin{align*}
\delta_2 = \arctan\left( {V_{2,y}/V_{2,x}} \right).
\end{align*}
In contrast, 1st harmonic signal is much weaker and it is often difficult to evaluate $\delta_1$ from $V_{1,x}$ and $V_{1,y}$. 
Thus, we use the fact that $\delta_n$ is related to $t_0$ as
\begin{align*}
    \delta_1 &= \pi - \omega t_0, \\
    \delta_2 &= \pi/2 - 2\omega t_0.
\end{align*}
From these equations, we obtain $\delta_1$ from $\delta_2$ as
\begin{align*}
    \delta_1 = \pi + \frac{\delta_2 - \pi/2}{2} = \frac{3}{4}\pi + \frac{\delta_2}{2}.
\end{align*}
Using these phase factors, we can accurately rotate the measured signals to obtain the in-phase signals $V_{\mathrm{in},n}$ as
\begin{align}
    V_{\mathrm{in},n} = 
        V_{x,n}\cos(\delta_n) + V_{y,n}\sin(\delta_n) 
        \label{eq: phase-corrected-in-phase-voltage}
\end{align}
Finally, we obtain the Kerr angle $\thetaK$ from the in-phase components as 
\begin{align}
 \theta\subm{tot} \simeq \tan(\thetaT) = \frac{1}{2}\frac{J_2(2\phim)}{J_1(2\phim)} \frac{V_{\mathrm{in},1}}{V_{\mathrm{in},2}}.
\end{align}
Typically we choose $2\phim = 1.84$, corresponding to the peak in $J_1(2\phi_m)$, so that the 1st component signal is maximized. 
This modulation depth yields $J_2(1.84)/J_1(1.84) =  0.543$.
Then, we reach the final relation
\begin{align*}
    \thetaT \simeq 0.271 \frac{V\subm{in,1}}{V\subm{in,2}}.
\end{align*}
Notice that, if we use voltage amplitudes without phase information, $V_{r,n} = \sqrt{ (V_{x,n})^2 + (V_{y,n} )^2}$, the resultant MO angle $\theta_{\mathrm{tot},r} \simeq 0.271 V_{r,1}/V_{r,2}$ would become always positive: i.e. the sign of the Kerr angle cannot be obtained without the phase-resolved analysis.

\subsection{Background subtraction}

It is inevitable that the lock-in amplifier reading contains small non-zero offsets.
Such voltage offsets can results in error in the Kerr angle, especially when voltage readings are small.
In the present work, we occasionally perform scan with zero optical output.
Then the reading of such blank scans should be solely attributable to the offsets.
We found that the offsets do not change in time. 
Thus, we determined the offset by averaging all the readings in one of the blank scans.
This evaluation yielded offsets in $V_{x,1}$ and $V_{y,1}$ as 0.716~$\mu$V and 5.44~$\mu$V, respectively.
These offsets are subtracted from the raw readings before the phase-resolved analysis
For $V_2$, since offsets are found to be negligibly  smaller than the $V_2$ readings, we did not perform offset subtraction.

We can also check the validity of this offset from the data in pixels outside the sample.
Since the Kerr angle should be zero outside the sample, the $V_1$ readings outside the sample should equal to the background.
We confirmed that the offsets determined from a blank scan and those from readings outside the sample are consistent with each other.

In addition, MOKE measurements in magnetic fields can be affected by background contribution due to the magneto-optical Faraday effect (change of state of polarization upon transmission through a magnetic material or a material in magnetic fields) of optic parts such as the lens and quarter waveplate. 
Nevertheless, the Faraday effect background is absent in the data shown in the present paper, since all measurements are performed in zero field. 

\subsection{Calibration of the sign of the magneto-optical angle}

Even with the careful phase-resolved analysis, there still are degrees of freedom that determine the sign of the obtained Kerr angle.
The sign can vary various factors such as the direction of the 1/4 waveplate with respect to the fast/slow axes of the PM fiber.
Thus, one needs a calibration on the sign of the Kerr angle using materials with known Kerr responses.

In this study, we performed scans of various ferromagnets placed on nickel-coated neodymium magnets.
For photon energy of 0.8~eV (wavelength of 1550~nm), nickel should have positive $\thetaK$ under positive magnetic fields, whereas cobalt is known to exhibits negative $\thetaK$~\cite{Delin1999.PhysRevB.60.14105}.
Thus, $\thetaK$ scans of cobalt placed on a nickel-coated neodymium magnets result in sign contrast between the sample and the magnet. 
We indeed observed that the nickel on the magnet shows positive $\thetaK$ and the sign is reversed on cobalt.
From these results, we can confirm that the sign of our apparatus is correct.

\subsection{Spatial resolution}

To experimentally verify the spatial resolution of our MOKE microscope, we investigate reflectivity mapping at edges of the sample. 
Figure~\ref{fig:sup:spotsize}\textbf{d} shows the position dependence of the 2nd harmonic voltage, which is proportional to the reflectivity, at several parts of the sample across its edge.
These data are taken from the mapping in Fig.~\ref{fig:sup:spotsize}{\bf a}
One can see that the profile exhibits a sharp decrease at the sample edge with less than 2~$\mu$m, indicating that our optics can resolve structures of such sizes. 

Assuming that the sample edge can be modeled as a step function and the optical spot has a gaussian profile with the standard deviation of $\sigma\subm{spot}$, the observed reflectivity curve should be a convolution of a step function and a gaussian function, namely the complementary error function $\mathrm{erfc}(x)$.
Then, by fitting the edge profile using the formula $V_{2r}(y) = V_0\, \mathrm{erfc}[(y-y_0)/\sigma\subm{spot}] + V\subm{offset}$, we can estimate $\sigma\subm{spot}$.
Results of such fittings are shown as the solid curves in Fig.~\ref{fig:sup:spotsize}\textbf{d}.
The obtained $\sigma\subm{spot}$ varies in the range 0.5 -- 2.4~$\mu$m. 
An error-weighted average yielded the spot size of 
$1.16~\pm 0.12$~$\mu$m. 
Thus, the spatial resolution of our MOKE microscope is around 1~$\mu$m.
Note that $\sigma\subm{spot}$ roughly corresponds to the half of the focus diameter, and thus these results agree with the specification of the focuser, whose focused spot diameter is 2.7~$\mu$m.

\begin{figure} 
	\centering
	\includegraphics[width=1.0\textwidth]{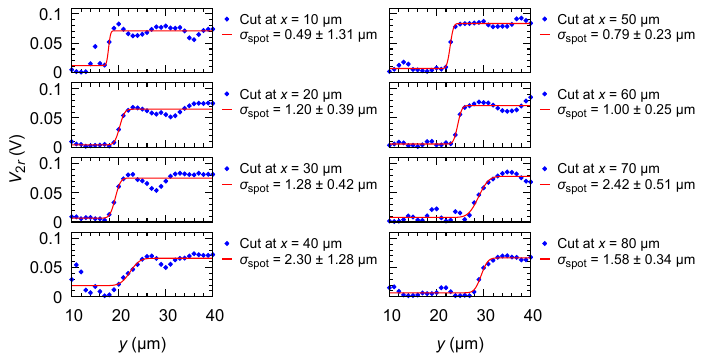} 
	\caption{
    \rev{\textbf{Investigations of spatial resolution of our scanning MOKE microscope.}
    Spatial resolution of our scanning MOKE microscope. The blue circles shows $y$ dependence of the 2nd harmonic signal, which is proportional to the reflectivity.
    The data points are extracted from \textbf{a}, at several $x$ positions specified in the legend.
    The red curves present results of fitting using the complementary error function with the width of $\sigma\subm{spot}$ with an offset.
    The fittings yield $\sigma\subm{spot} = 0.5$ -- 2.4~$\mu$m, whose error-weighted average is $1.16 \pm 0.12$~$\mu$m, demonstrating that our special resolution of around 1~$\mu$m.
    Considering the fact that $\sigma\subm{spot}$ should roughly correspond to the half of the light spot diameter, the results are consistent with the spot diameter of 2.7~$\mu$m.
	}	}
	\label{fig:sup:spotsize} 
\end{figure}

\clearpage 

\section{Details of the theoretical calculation}

\begin{figure}[h] 
	\centering
    \includegraphics[width=0.8\textwidth]{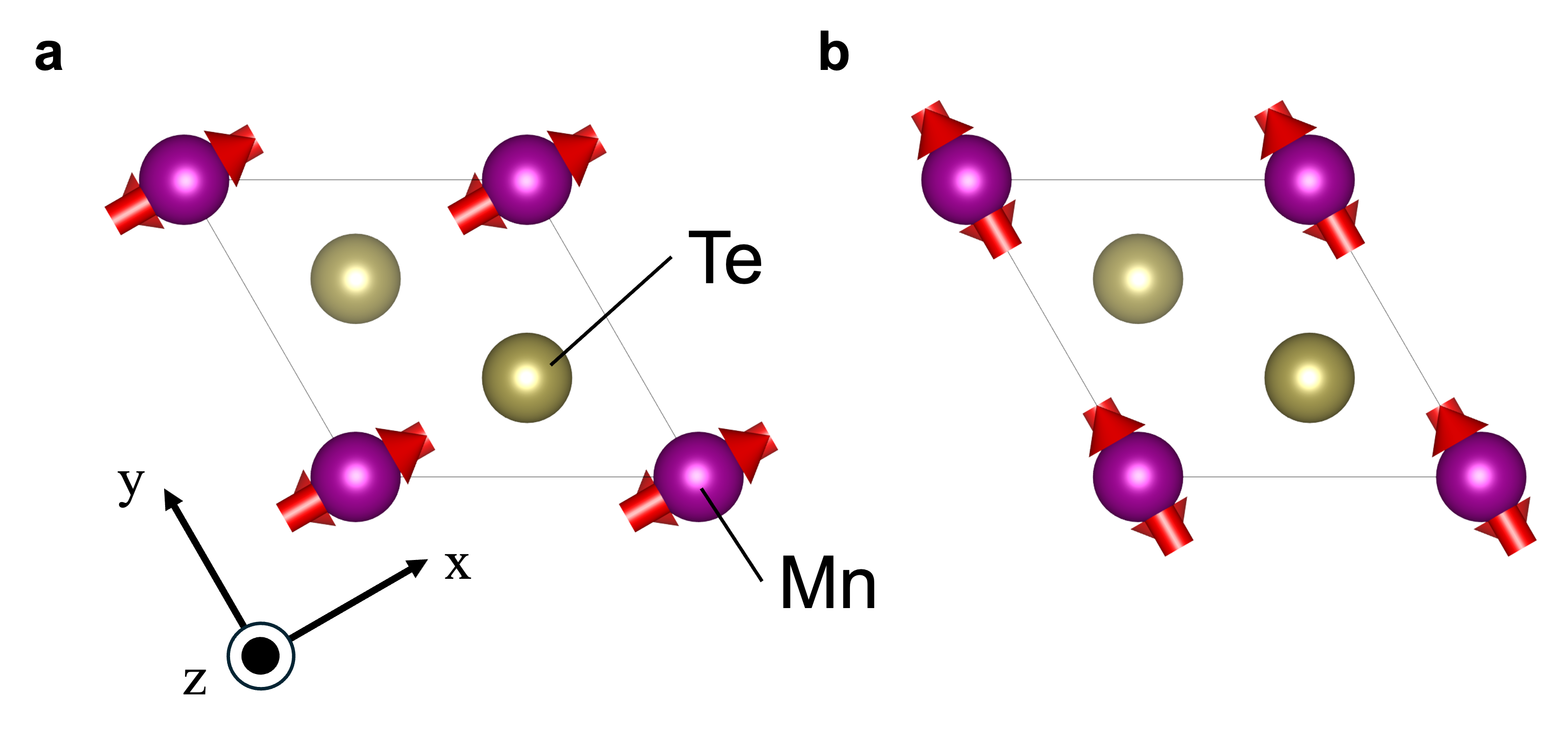}
	\caption{\textbf{Magnetic structures considered in the calculations.}
    (\textbf{a}) Magnetic structure with $m'm'm$ symmetry, corresponding to experiment.
    (\textbf{b}) Magnetic structure with $mmm$ symmetry.
	}
	\label{fig:sup:magnetic_structure}
\end{figure}

\begin{figure} 
	\centering
    \includegraphics[width=1.0\textwidth]{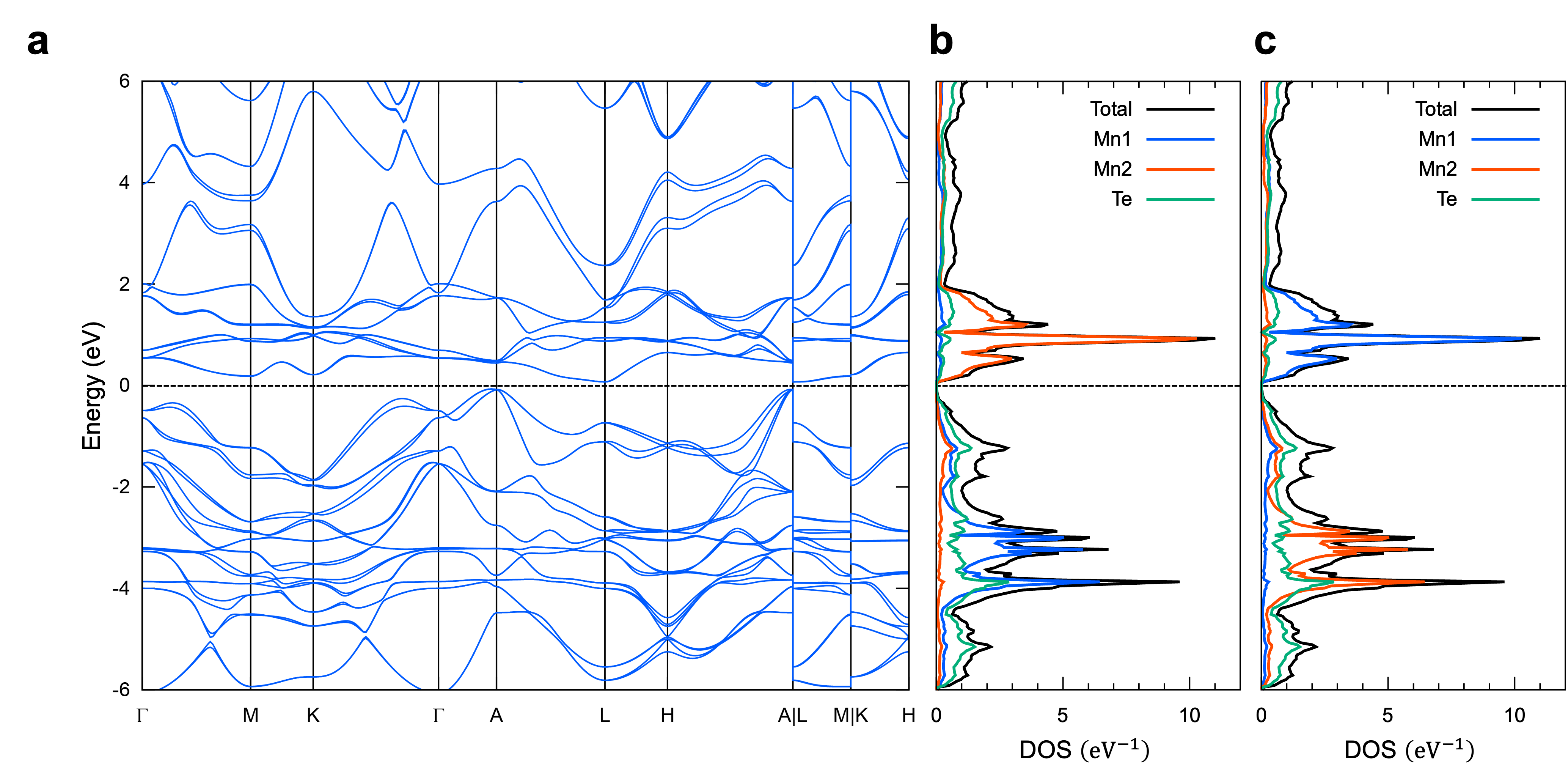} 
	\caption{\textbf{Electronic structure in the antiferromagnetic state with the $m'm'm$ structure.}
    (\textbf{a}) Energy dispersion.
    (\textbf{b}--\textbf{c}) Spin-resolved density of states. Mn1 and Mn2 represent two inequivalent Mn sites.
	}
	\label{fig:sup:band}
\end{figure}

\rev{Two magnetic structures considered in the present calculations are shown in Fig.~\ref{fig:sup:magnetic_structure}. 
The $m'm'm$ magnetic structure corresponds to the experiment, whereas the $mmm$ structure is also examined to clarify the origin of the Kerr effect.
The calculations were performed based on the density functional theory (DFT) within the full potential local orbital (FPLO) method~\cite{FPLO}, using the generalized gradient approximation (GGA) under full relativistic condition.
The electronic band structure used for the calculation is shown in Fig.~\ref{fig:sup:band}.
}

\clearpage

\section{Domain wall and sub-domain structures}
\label{sec:domain-wall}

To investigate the fine structure in more detail, we performed 1-$\mu$m-step scans within altermagnetic domains.
A representative scan result is shown in Fig.~\ref{fig:sup:spotsize}\textbf{a}--\textbf{c}.
In this figure, sharp domain walls between positive and negative $\thetaK$ was observed.
From line-cuts around a sharp domain wall (the bottom panel of \textbf{c}), the observed widths of the domain walls can be estimated by fitting $\thetaK(x) = \theta_1 \mathrm{erf}[2(x-x_0)/w\subm{DW}] + \theta_0$ to the line profile, where $\mathrm{erf}(x)$ is the error function, $\theta_1$ and $\theta_0$ are coefficients, $x_0$ is the domain-wall position, and $w\subm{DW}$ is the domain-wall width. 
The fitting result, shown with the red dashed curve, yields $w\subm{DW} = 2.4\pm 1.2$~$\mu$m.
This observed width is very close to the spatial resolution of our setup (Fig.~\ref{fig:sup:spotsize}\textbf{d}). 
Thus the actual domain-wall width is very likely to be much smaller. 
We comment that domain-wall width of 0.12 -- 0.13~nm in thin films~\cite{Amin2024.Nature.636.348} and 0.39~nm in single crystal flakes~\cite{Yamamoto2025.PhysRevAppl.24.034037} has been reported using X-ray techniques.

Within each domain, we find bubble-like structures with modulation of $\thetaK$ of the typical sizes of a few $\mu$m.
This observation of sub-domain structure is quite interesting, 
since ordinary magnetic domains should have homogeneous magnetic structures within a domain.
We have checked that the sub-domain structures do not have strong correlation with the surface reflectivity, as seen in the comparison of the panels in Fig.~\ref{fig:sup:spotsize}\textbf{c}.
Although their origins are not yet understood,
these in-domain structures partly resemble the microscopic domains observed in thin films using X-ray techniques~\cite{Amin2024.Nature.636.348}.
Possible origins of this sub-domain structure is discussed in the main text. 

\begin{figure} 
	\centering
	\includegraphics[width=1.0\textwidth]{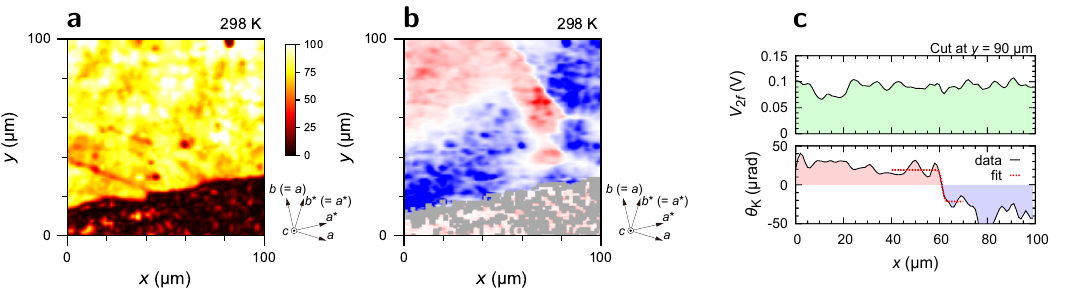} 
	\caption{\rev{\textbf{Mesoscopic domain structures.}
    \textbf{a}. 1-$\mu$m-step $V_{2f}$ mapping (reflectivity mapping) near the edge of the MnTe sample.
    \textbf{b}. Kerr mapping on the area same as \textbf{a}.
    \textbf{c}. Line profile of $V_{2F}(x)$ (top) and $\thetaK(x)$ (bottom) at $y=90$~$\mu$m. 
    There is no strong correlation between the surface reflectivity and the Kerr angle.
    The red dashed curve shows a result of fitting to the $\thetaK(x)$ near the domain wall.
    This fit yields the observed domain wall width of $w\subm{DW} \sim 2.4$~$\mu$m, which is likely to be limited by the apparatus spatial resolution. }
		}
	\label{fig:sup:spotsize} 
\end{figure}

\end{document}